# Real-time PCG Anomaly Detection by Adaptive 1D Convolutional Neural Networks

Serkan Kiranyaz, Morteza Zabihi, Ali Bahrami Rad, Anas Tahir, Turker Ince, Ridha Hamila, and Moncef Gabbouj

*Abstract*— The heart sound signals (Phonocardiogram - PCG) enable the earliest monitoring to detect a potential cardiovascular pathology and have recently become a crucial tool as a diagnostic test in outpatient monitoring to assess heart hemodynamic status. The need for an automated and accurate anomaly detection method for PCG has thus become imminent. To determine the state-of-the-art PCG classification algorithm, 48 international teams competed in the PhysioNet (CinC) Challenge at 2016 over the largest benchmark dataset with 3126 records with the classification outputs, normal (N), abnormal (A) and unsure – too noisy (U). In this study, our aim is to push this frontier further; however, we focus deliberately on the anomaly detection problem while assuming a reasonably high Signal-to-Noise Ratio (SNR) on the records. By using 1D Convolutional Neural Networks trained with a novel data purification approach, we aim to achieve the highest detection performance and a real-time processing ability with significantly lower delay and computational complexity. The experimental results over the high-quality subset of the same benchmark dataset shows that the proposed approach achieves both objectives. Furthermore, our findings reveal the fact that further improvements indeed require a personalized (patient-specific) approach to avoid major drawbacks of a global PCG classification approach.

## I. INTRODUCTION

According to World Health Organization (WHO) report in 2015, an estimated 17.7 million people died from cardiovascular diseases only in 2015, representing 37% of all premature deaths worldwide. The most common, and possibly the cheapest and the earliest clinical examination is heart auscultation, which can reveal several cardiac anomalies, such as certain arrhythmias, failures, murmurs, as well as diseases such as ventricular septal defects, and stenosis in aorta (McConnell, 2008) and many more. However, there are several limitations and practical problems in the human auditory system when it comes to phonocardiogram (PCG) signal analysis, despite the cognitive skills and expertise of the medical examiner. This is worsened by surrounding environmental noise and possibly large signal variations in the local recording areas. This brought the need for an automated, cost-effective and robust anomaly detection method for PCG signals. During the last 20 years several approaches (Guptaaet al., 2007; Balili et al. 2015; Moukadem et al., 2013; Zhang et al. 2011; Kao et al., 2011) have been proposed in the literature. However, most of them were tested on limited PCG datasets usually containing only a small number of records, e.g. a few dozens. Most of them used a particular machine learning approach and it is a well-known fact that they can be easily tuned to maximize the classification performance on a limited test set. Moreover, it is not feasible to perform any comparative evaluation between them due to the variations of the train and test datasets used. The need for a benchmark dataset which contains a sufficiently large collection of PCG records was imminent, and finally, the PhysioNet community (Physionet, 2016) composed the largest PCG dataset that was used as the benchmark dataset in PhysioNet (CinC) Challenge 2016, where 48 proposals have recently competed. The benchmark dataset contains 3126 records; however, some records were too difficult to analyze due to the significant noise; therefore, the classification outputs as, normal (N), abnormal (A) and noisy (U) were recommended by the Challenge. As a result, the objective of the competition is not only to detect record with anomalies (A) but also to detect those records which are "too noisy" (U). This was, therefore, a blended objective rather than a sole anomaly detection.

Our earlier work, Zabihi et al. (2016), was the 2[nd] ranked method in the PhysioNet (CinC) Challenge, with respect to this combined objective, was indeed the best-performing method for PCG anomaly detection. So taking contribution Zabihi et al. (2016) of the challenge as the reference, the three main objectives of this study are to: 1) further improve the state-of-the-art anomaly detection performance 2) reduce the false alarms (misclassification of normal records as abnormal), and 3) achieve a real-time processing ability with significantly lower delay and computational complexity. The detection of low-quality (too noisy) records is beyond the scope of this study and therefore, we shall assume that each PCG records has a sufficiently high Signal-to-Noise Ratio (SNR) and can be segmented with a reasonably high accuracy. SNR is defined as the ratio of the signal power to noise power. Specifically, $\text{SNR} = 10\log\left(\sigma_{signal}^2 / \sigma_{noise}^2\right)$. The latter assumption has already been fulfilled by several recent methods (i.e., >95% accuracy is achieved in Moukadem et al., 2013. and Sattar et al., 2015). The current state-of-the-art method in Zabihi et al. (2016) is also immune to PCG segmentation errors since it classifies the entire PCG record as a whole. However, the method

S. Kiranyaz is with Electrical Engineering Department, College of Engineering, Qatar University, Qatar; e-mail: mkiranyaz@qu.edu.qa.
M. Zabihi is with with the Department of Signal Processing, Tampere University of Technology, Finland; e-mail: Morteza.Zabihi@tut.fi.
A. B. Rad is with the Department of Electrical Engineering and Automation, Aalto University, Espoo, Finland; email: abahramir@gmail.com.
A. Tahir is with Electrical Engineering Department, College of Engineering, Qatar University, Qatar; e-mail: at1003225@student.qu.edu.qa .

T. Ince is with the Electrical & Electronics Engineering Department, Izmir University of Economics, Turkey; e-mail: turker.ince@izmirekonomi.edu.tr .
R. Hamila is with Electrical Engineering Department, College of Engineering, Qatar University, Qatar; e-mail: hamila@qu.edu.qa.
M. Gabbouj is with the Department of Signal Processing, Tampere University of Technology, Finland; e-mail: Moncef.gabbouj@tut.fi.

also presents a high computational complexity due to the various hand-crafted feature extraction techniques employed. Hence the real-time processing may not be feasible. not only because of the such high computational complexity, but also due to the fact that the entire PCG record should be acquired first.

In order to address these drawbacks, in this article we propose a novel PCG anomaly detection scheme that employs an adaptive 1D Convolutional Neural Network (CNN) and a data purification process that is embedded into the Back-Propagation (BP) training algorithm. "Deep" CNNs are feed-forward artificial neural networks which were initially developed as the crude models of mammalian visual cortex. Deep CNNs have recently become the *de-facto* standard for many visual recognition applications (e.g., object recognition, segmentation, tracking, etc.) as they achieved the state-of-the-art performance (Ciresan et al., 2010; Scherer et al. 2010; Krizhevsky et al., 2012) with a significant performance gap. Recently, 1D CNNs have been proposed for pattern recognition for 1D signals such as patient-specific ECG classification (Kiranyaz et al., 2015 and 2017), structural health monitoring (Abdeljaber, Avci, et al., 2018; Abdeljaber et al., 2017; Avci et al., 2018) and mechanical and motor fault detection systems, (Eren et al., 2018; Abdeljaber, Sassi, et al., 2018; Ince et al., 2016). Because there are numerous advantages of using an adaptive and compact 1D CNN instead of a conventional (2D) deep counterpart. First of all, compact 1D CNNs can be efficiently trained with a limited dataset in 1D (e.g., Kiranyaz et al., 2015 and 2017; Abdeljaber, Avci, et al., 2018; Abdeljaber et al., 2017; Avci et al., 2018; Eren et al., 2018; Abdeljaber, Sassi, et al., 2018; Ince et al., 2016) while the 2D deep CNNs, besides the 1D to 2D data transformation, require datasets with massive size, e.g., in the "Big Data" scale in order to prevent "Overfitting" problem. In fact, this requirement alone makes deep 2D CNNs inapplicable to many practical problems that have limited datasets including the problem addressed in this study. The crucial advantage of the CNNs is that both feature extraction and classification operations are fused into a single machine learning body to be jointly optimized to maximize the classification performance. This eliminates the need for hand-crafted features or any other post processing. So this is the key property to achieve the aforementioned objectives. Furthermore, due to the simple structure of the 1D CNNs that requires only 1D convolutions (scalar multiplications and additions), a real-time and low-cost hardware implementation of the proposed approach is quite feasible. Finally, it is a well-known fact that the abnormal PCG records may and usually do have certain amount of normal sound beats and hence they will inevitably cause a certain level of learning confusion which eventually leads to a degradation on the classification performance. The data purification that is embedded into the BP training is designed to reduce this during the training phase (offline) without causing any further time delay on the detection.

The rest of the paper is organized as follows: Section II first presents the previous work in PCG classification with the main challenges tackled. Then PhysioNet/CinC Challenge will be presented and the focus is particularly drawn on the best method (Zabihi et al., 2016) of the challenge that achieves the highest score in PCG anomaly detection. The proposed anomaly detection approach using an adaptive 1D CNN is presented in Section III. In Section IV, extensive comparative evaluations against Zabihi et al. (2016) will be performed using the standard performance metrics over the two datasets each of which is created from the PhysioNet/CinC benchmark database. Finally, Section V concludes the paper and suggests topics for future research.

## II. PRIOR WORK

### A. Overview

Heart auscultation is performed on the anterior chest wall and used as the primary diagnostic method to evaluate the heart function. A PCG signal shows the auscultation of turbulent blood flow and the timing of the heart valves' movements. The main components of a PCG signal in each heartbeat are S1 (first) and S2 (second) sounds, which are related to the closure of mitral and tricuspid valves, and the closure of aortic and pulmonic valves, respectively.

Hearing abnormal heart sounds, such as transients and murmurs, in a PCG signal is usually considered pathological. These abnormal sounds can provide information about the blockage of the valve when it is open (stenosis), the leakage of the valve when it is closed (insufficiency), the increase of blood flow, etc. However, the correct interpretation of such signals depends on the acuity of hearing and the experience of the observer.

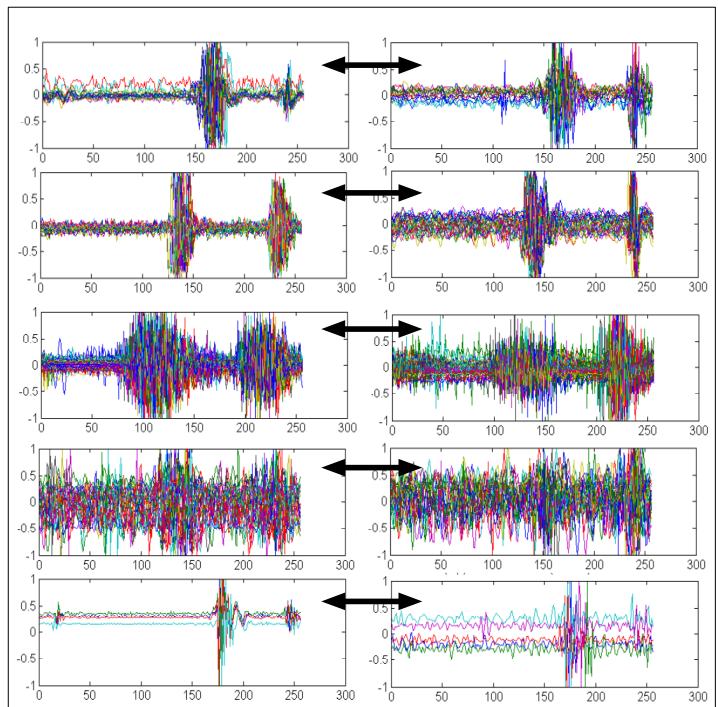

**Figure 1: N (left) vs. A (right) beats from different subject records in Physionet/CinC dataset (Physionet, 2016).**

Several studies have been presented to automatically identify PCG anomalies using signal processing and machine learning techniques. There are two types of PCG signal analysis and anomaly detection methods. The first type analyzes the entire PCG record without segmentation (Yuenyong et al., 2011; Deng et al., 2016) whereas the second type uses temporal beat segmentation, i.e. identifying the cardiac cycles and localizing the position of the first (S1; beginning of the systole) and second (S2; end of the systole) primary heart sounds. Certain variations over S1 and S2 properties such as their duration or intensities can be considered as the primal signs of cardiac anomalies. For PCG segmentation (or beat detection) there exist various prior works using different envelope extraction methods such as Shannon energy (Guptaa et al., 2007),

Shannon entropy (Moukadem et al., 2013), Hilbert-Huang transform (Zhang et al., 2011), and autocorrelation (Kao et al., 2011). Some methods use envelope extraction based on wavelet transform to gain the frequency characteristics of S1 and S2 sound (Huiying et al., 1997).

PCG classification is especially challenging due to high variations among the N and A type PCG patterns and the noise levels. Figure 1 shows some typical PCG beats where different records in the benchmark Physionet/CinC dataset (Liu et al., 2016) have entirely different N beats, which may, however, show high level of structural similarities to other A beats (e.g. see the pairs with arrows in the figure). Obviously, past approaches that relied on only one or few hand-crafted features may not characterize all such inter- and intra-class variations and thus they have usually exhibited a common drawback of having an inconsistent performance when, for instance, classifying a new patient's PCG signal. This makes them unreliable for wide clinical use.

### B. PhysioNet/CinC Challenge

For this challenge, 3126 PCG labelled records (including training and validation sets) were provided by Physionet/Computing in Cardiology Challenge 2016 (Physionet, 2016). The provided database was sourced by several international contributors. The database collected from both healthy and pathological subjects in clinical and nonclinical environments. In the database, each PCG record is labeled as N or A. In addition, the quality of each recording was also identified either as good or bad quality. The task was to design a model that can automatically classify the PCG recordings as N, A, or too noisy to evaluate. More detailed information can also be found in Liu et al. (2016).

### C. The Best Anomaly Detection Method Zabihi et al. (2016)

As illustrated in , a set of 18 features was selected from a comprehensive set of hand-crafted features using a greedy wrapper-based feature selection algorithm. The selected features consisted of the linear predictive coefficients, natural and Tsallis entropies, mel frequency cepstral coefficients, wavelet transform coefficients, and power spectral density functions (Zabihi et al., 2016). Then, the features were classified through a two-step algorithm. In the first step, the quality of PCG recordings was detected, and in the second step, the PCG records with good quality were further classified into N or A classes.

The classification algorithm in Zabihi et al. (2016) consists of an ensemble of 20 feedforward sigmoidal Neural Networks (NNs) with two hidden layers, and 25 hidden neurons in each layer. The output layer has 4 neurons to detect the signal quality (good vs. bad) and anomaly (N vs. A), simultaneously. This ensemble of NNs was trained via a 20-fold cross-validation committee with Bayesian regularization back-propagation training algorithm (MacKay, 1992), and a bootstrap resampling method to make the size of the N and A classes balanced. The final classification rule which was learned via a 10-fold cross-validation method, combines the decision of all 20 NNs such that if at least 17 NNs classified a signal as bad quality, the algorithm recognized it as bad quality, otherwise it is detected as good quality. Good quality signals were further classified as A if at least 7 NNs recognized it as A, otherwise, it was detected as N.

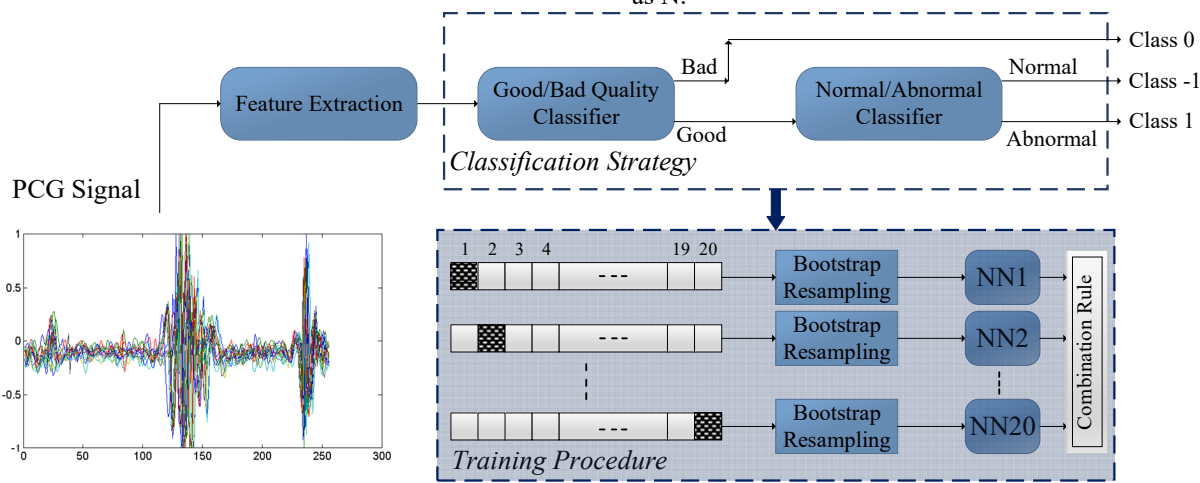

**Figure 2: Schematic demonstration of classification strategy and training procedure of Zabihi et al. (2016).**

Since the objective function (i.e., the evaluation mechanism) and the dataset of the PhysioNet (CinC) Challenge were different from the current study, we adapted the method in Zabihi et al. (2016), i.e., in the output layer of ANNs we now used two neurons rather than four since we do not need to detect the quality of PCG signals anymore.

### III. THE PROPOSED APPROACH

Here we proposed a novel approach for real-time monitoring heart sound signals for robust and accurate anomaly detection. To accomplish the aforementioned objectives, we used an adaptive 1D CNN at the core of the system that is trained by labelled and segmented PCG records with sufficiently high SNR. Data purification is periodically performed during the BP training in order to reduce the confusion due to the potential normal (N) beats in abnormal (A) records. Once the 1D CNN is properly trained, it can classify each monitored PCG beat in real-time, or alternatively classify a PCG record of a patient as N or A.

As illustrated in Figure 3, the PCG signal can be acquired either from a PCG monitoring device in real-time or retrieved from a PCG file. The data processing block first segments the PCG stream into PCG beats. Each PCG beat, $p$, is linearly normalized in the range of [-1, 1] as follows:

$$p_i = 2 \frac{p_i - \min(p)}{\max(p) - \min(p)} - 1 \qquad (1)$$

where $p_i$ is the $i^{th}$ sample in the PCG beat, $p$. Eq. (1) linearly maps the minimum and maximum samples to -1 and 1, respectively, and hence negates the effect of the sample amplitudes in the learning process. Each normalized beat is then fed into the 1D CNN classifier that has already been trained as shown in Figure 4. For a real-time operation the output of the 1D CNN can be used directly or alternatively can be processed by a majority rule to obtain the final class decision of the entire stream, e.g., the stream is classified as A if more than 25% of the beats are abnormal. This threshold should be determined in a practical way so that it should be sufficiently high to prevent false-alarms due to classification noise while not too high to detect those abnormal records with few abnormal PCG beats. Therefore, the "Final Decision" block determines the final class type of a PCG file while its utilization is optional for real-time PCG anomaly detection.

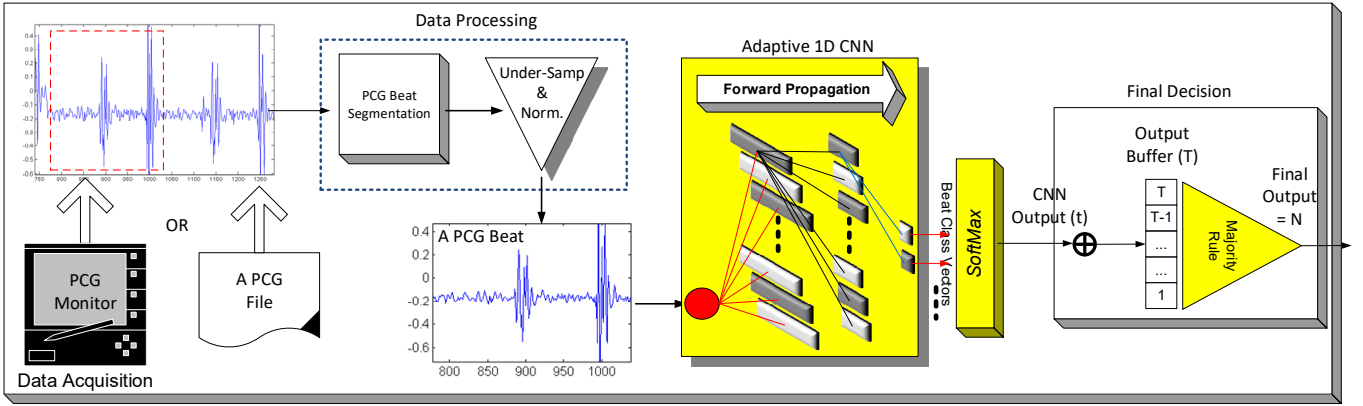

**Figure 3: The proposed system architecture for real-time PCG anomaly detection after the BP training of the 1D CNN as illustrated in Figure 4.**

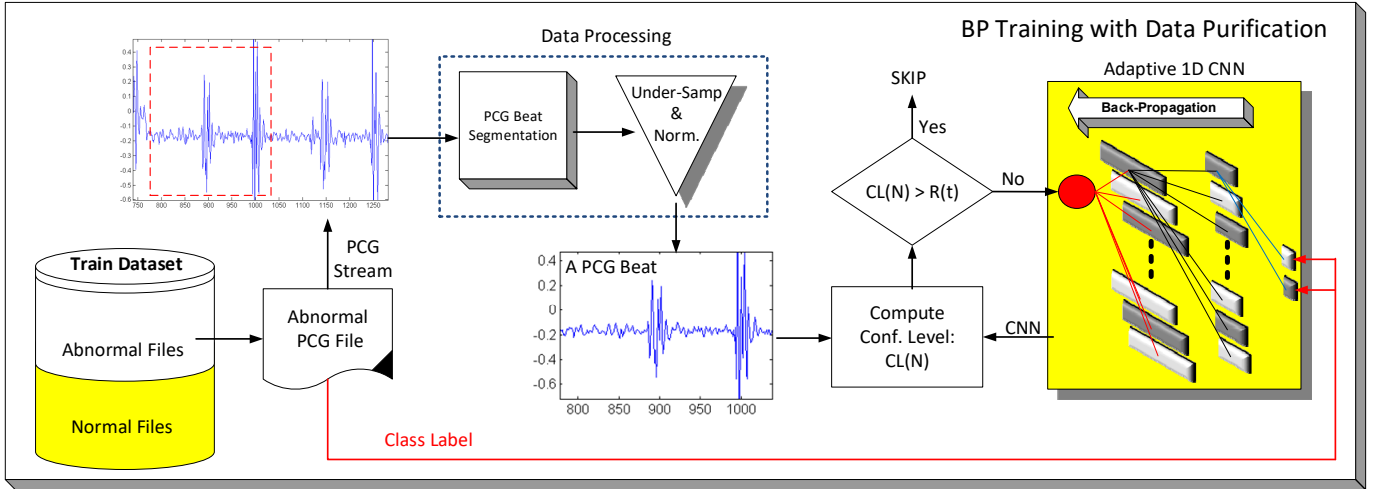

**Figure 4: The (offline) BP training of the adaptive 1D CNN for abnormal PCG beats.**

In the rest of this section, an overview of 1D CNNs will first be presented and the BP training with data purification will be detailed next.

*A. Adaptive 1D CNN Overview*

Convolutional Neural Networks (CNNs) are feed-forward artificial neural networks and they are mainly used for 2D signals such as images and video (Ciresan et al., 2010; Scherer et al., 2010; Krizhevsky et al., 2012). In this study, we use compact and adaptive 1D CNN configuration in order to accomplish the objectives presented in Section I. Compact 1D CNNs have a simple and shallow configuration with only few hidden layers and neurons. Typically, number of parameters in the network is only few hundreds and the network can adapt itself to the duration (resolution) of each PCG segment.

There are two types of layers in the adaptive 1D CNNs: 1) CNN-layers where both 1D convolutions and sub-sampling occur, and 2) Fully-connected layers that are identical to the hidden and output layers of a typical Multi Layer Perceptron (MLP). As illustrated in Figure 5, the CNN-layers decompose the raw data in several scales and learn to extract such features that can be used by the fully connected layers (or MLP-layers) for classification. Therefore, both feature extraction and classification operations are fused into one body that can be optimized to maximize the classification performance. In the CNN-layers, the 1D forward propagation (FP) between the two CNN layer can be expressed as,

$$x_k^l = b_k^l + \sum_{i=1}^{N_{l-1}} conv1D(w_{ik}^{l-1}, s_i^{l-1}) \qquad (2)$$

where $x_k^l$ is the input, $b_k^l$ is the bias of the $k^{th}$ neuron at layer $l$, and $s_i^{l-1}$ is the output of the $i^{th}$ neuron at layer $l$-1. $w_{ik}^{l-1}$ is the kernel from the the $i^{th}$ neuron at layer $l$-1 to the $k^{th}$ neuron at layer $l$.

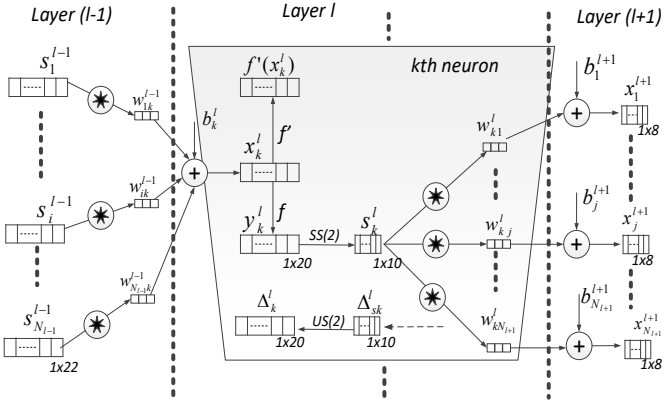

**Figure 5: Three CNN layers of a 1D CNN.**

With such an "adaptive" design it is aimed that the number of hidden CNN layers can be set to any practical number because the sub-sampling factor of the output CNN layer (the hidden CNN layer just before the first MLP layer) is set to the dimensions of its input map, e.g., in Figure 5 if the layer l+1 would be the output CNN layer, then the sub-sampling factors for that layer is automatically set to $ss = 8$ since the input map dimension is 8 in this sample illustration. Besides the sub-sampling, note that the dimension of the input maps will gradually decrease due to the convolution without zero padding, i.e., in Figure 5, the dimension of the neuron output is 22 at the layer $l$-1 that is reduced to 20 at the layer $l$. As a result of this, the dimension of the input maps of the current layer is reduced by $K$-1 where $K$ is the size of the kernel.

### B. Back-Propagation with Data Purification

We shall now briefly formulate the back-propagation (BP) steps while skipping the detailed derivations as further details can be found in Kiranyaz et al. (2015).

The BP of the error starts from the MLP output layer. Let $l$=1 and $l=L$ be the input and output layers, respectively. The error in the output layer can be written as,

$$E = E(y_1^L, y_2^L) = \sum_{i=1}^{2}(y_i^L - t_i)^2 \quad (3)$$

For an input vector $p$, and its corresponding output and target vectors, $[y_1^L, y_2^L]$ and $[t_1^L, t_2^L]$, respectively, we are interested to find out the derivative of this error with respect to an individual weight (connected to that neuron, $k$) $w_{ik}^{l-1}$, and bias of the neuron $k$, $b_k^l$, so that we can perform gradient descent method to minimize the error accordingly. Once all the delta errors in each MLP layer are determined by the BP, then weights and bias of each neuron can be updated by the gradient descent method. Specifically, the delta error of the $k^{th}$ neuron at layer $l$, $\Delta_k^l$, will be used to update the bias of that neuron and all weights of the neurons in the previous layer connected to that neuron, as:

$$\frac{\partial E}{\partial w_{ik}^{l-1}} = \Delta_k^l y_i^{l-1} \quad and \quad \frac{\partial E}{\partial b_k^l} = \Delta_k^l \quad (4)$$

So from the input MLP layer to the output CNN layer, the regular (scalar) BP is simply performed as,

$$\frac{\partial E}{\partial s_k^l} = \Delta s_k^l = \sum_{i=1}^{N_{l+1}} \frac{\partial E}{\partial x_i^{l+1}} \frac{\partial x_i^{l+1}}{\partial s_k^l} = \sum_{i=1}^{N_{l+1}} \Delta_i^{l+1} w_{ki}^l \quad (5)$$

Once the first BP is performed from the next layer, $l$+1, to the current layer, $l$, then we can further back-propagate it to the input delta, $\Delta_k^l$. Let zero order up-sampled map be: $us_k^l = up(s_k^l)$, then one can write:

$$\Delta_k^l = \frac{\partial E}{\partial y_k^l} \frac{\partial y_k^l}{\partial x_k^l} = \frac{\partial E}{\partial us_k^l} \frac{\partial us_k^l}{\partial y_k^l} f'(x_k^l) = up(\Delta s_k^l)\beta f'(x_k^l) \quad (6)$$

where $\beta = (ss)^{-1}$ since each element of $s_k^l$ was obtained by averaging $ss$ number of elements of the intermediate output, $y_k^l$. The inter-BP (among CNN layers) of the delta error ($\Delta s_k^l \xleftarrow{\Sigma} \Delta_i^{l+1}$) can be expressed as,

$$\Delta s_k^l = \sum_{i=1}^{N_{l+1}} conv1Dz\left(\Delta_i^{l+1}, rev(w_{ki}^l)\right) \quad (7)$$

where $rev(.)$ reverses the array and $conv1Dz(.,.)$ performs full convolution in 1D with $K$-1 zero padding. Finally, the weight and bias sensitivities can be expressed as,

$$\frac{\partial E}{\partial w_{ki}^l} = conv1D(s_k^l, \Delta_i^{l+1}) \quad \frac{\partial E}{\partial b_k^l} = \sum_n \Delta_k^l(n) \quad (8)$$

#### 1) Back-Propagation Flowchart

As a result, the iterative flow of the BP for the N beats in normal PCG records in the training set can be stated as follows:

1) Initialize weights and biases (usually randomly, U(-a, a)) of the network.
2) For each BP iteration DO:
   a. For each PCG beat in the dataset, DO:
      i. **FP**: Forward propagate from the input layer to the output layer to find outputs of each neuron at each layer, $y_i^l, \forall i \in [1, N_l] \, and \, \forall l \in [1, L]$ .
      ii. **BP**: Compute delta error at the output layer and back-propagate it to first hidden layer to compute the delta errors, $\Delta_k^l, \forall k \in [1, N_l] \, and \, \forall l \in [2, L-1]$ .
      iii. **PP**: Post-process to compute the weight and bias sensitivities using Eq. (8)
      iv. **Update**: Update the weights and biases with the (cumulation of) sensitivities found in (c) scaled with the learning factor, $\varepsilon$:

$$w_{ik}^{l-1}(t+1) = w_{ik}^{l-1}(t) - \varepsilon \frac{\partial E}{\partial w_{ik}^{l-1}}$$
$$b_k^l(t+1) = b_k^l(t) - \varepsilon \frac{\partial E}{\partial b_k^l}$$
(9)

*2) Data Purification*

The proposed method works over the PCG beats (frames) while the labelling information is for the PCG records (files) of the benchmark dataset created by the Cinc competition. So when a PCG record is labelled as N, it is clear that all the PCG beats in this record are normal. However, if a PCG record is labelled as A, this does not necessarily mean that all the PCG beats are abnormal. In this case, there are certain amount of normal beats in this abnormal record and it is quite possible that these normal beats can be even the majority of this record. Therefore, while training the 1D CNN, these normal beats should not be included in the training with a label "A" because 1D CNN will obviously "confuse" while processing a normal beat having the label "A". To avoid this confusion we altered the conventional BP process for back-propagating the error from each abnormal PCG beat from an abnormal file. As illustrated in Figure 4, during each BP iteration, the error from a PCG beat in an abnormal record will be back-propagated if the current CNN does not classify it as a N beat with a high confidence level. In other words, if the current CNN classifies it as a N beat with a sufficiently high confidence level, i.e., CL(N) > R(t) where t is the current BP iteration, then it will be dropped out of the training process until the next periodic check-up. The threshold, R(t), for the confidence level can also be periodically reduced since the CNN can distinguish the N beats with a higher accuracy as the BP training advances and thus during the advance stages of the BP those beats that are classified as N with a reasonable confidence level can now be conveniently skipped from the training. Note that the aim of this data purification is to reduce a potential confusion (e.g., N beats in an abnormal PCG record). An alternative action would be to reverse their truth labels (A) to N and still include them in the BP training. However, this is only an "expected" outcome indicated by a partially trained CNN, not a proven one. Hence such an approach may cause a possible error accumulation. Even if some of those beats are still abnormal despite the CNN classification, skipping them from the training may only reduce the generalization capability of the classifier due to reduced training data.

Accordingly, the **FP** and **BP** operations in the 2$^{nd}$ BP step can be modified as follows:

2) For each BP iteration, t, DO:
   a) For each PCG beat in an abnormal record, DO:
     i. **FP**: Forward propagate from the input layer to the output layer to find outputs of each neuron at each layer, $y_i^l, \forall i \in [1, N_l] \text{ and } \forall l \in [1, L]$.
     ii. **Periodic Check (Pc):** At every Pc iteration DO:
      **Compute**: CL(N) = $50(y_1^L - y_2^L)$. If CL(N) > R(t), mark this beat as "skip".
     iii. **Skip**: If the beat is marked as "skip", drop this beat out of BP training and proceed with the next beat.
     iv. **BP**: Compute delta error at the output layer and back-propagate it to first hidden layer to compute the delta errors, $\Delta_k^l, \forall k \in [1, N_l] \text{ and } \forall l \in [2, L-1]$.

## IV. EXPERIMENTAL RESULTS

In this section, the experimental setup for the test and evaluation of the proposed systematic approach for PCG anomaly detection will first be presented. Then an evaluation of the data purification process will be visually performed over two PCG sample records and the results will be discussed. Next using the conventional performance metrics, the overall abnormal beat detection performance and the robustness of the proposed system against variations of the system and network parameters will be evaluated. Finally, the computational complexity of the proposed method for both offline and real-time operations will be reported in detail.

### A. Experimental Setup

In the PCG dataset created in the Challenge, the sampling rate is 4Khz. In a normal rhythm, one can expect around 1 beat per second. In a paced activity, this can go up to maximum 4 beats/second. So the number of samples in a beat will vary between 1000 – 4000 samples. In order to feed the raw PCG signal as the input frame to 1D CNN, the frame size should be fixed to constant value. Therefore, in this study each segmented PCG beat has down-sampled to 1000 samples, which is a practically high value to capture each beat with a sufficiently high resolution and allows the network to analyze the beat characteristics independent from its duration. We used a compact 1D CNN in all experiments with only 3 hideen CNN layers and 2 hidden MLP layers, in order to achieve the utmost computational efficiency for both training and particularly for real-time anomaly detection. The 1D CNN used in all experiments has 24, neurons in all (hidden) CNN and MLP layers. The output (MLP) layer size is 2 as this is a binary classification problem and the input (CNN) layer size is 1. The kernel size of the CNN is 41 and the sub-sampling factor is 4. Accordingly, the sub-sampling factor for the last CNN-layer is automatically set to 10. Figure 6 illustrates the compact network configuration used in the experiments.

In order to form the PCG dataset with a sufficiently high SNR, we estimated the noise variance of each record over the low signal activity (diastole) section of each PCG beat (e.g. the first 20% of each segment). Because when the signal activity is minimal, the variance computation will belong to noise and it will hence be inversely proportional to the SNR. To compute the noise average of a PCG record we will then take the average of the beat variances. According to their average noise variances, we then sorted the PCG records in the database and then took the top 1200 PCG records (900 N and 300 A) to constitute the high-SNR PCG dataset. The estimated average SNR of the worst (bottom-ranked) PCG record in this list is around -6 dB, which means that the noise power (variance) is 4 times higher than the signal power. In order to evaluate the noise effect over the proposed approach, we also form the 2$^{nd}$ dataset with the lowest possible SNR. For this purpose, we selected 1008 (756 N and 252 A) PCG records from the bottom of the sorted list. The estimated average SNR of the best (top-ranked) PCG record in this list is around -9 dB, which means that the noise power (variance) is 8 times higher than the signal power. Over both high- and low-SNR datasets, we performed 4-fold cross validation and for each fold, one partition (25%) is kept for testing while the other three (75%) are used for training. This allows us to test the proposed system over all the records. Since the BP training algorithm is a gradient descent algorithm, which has a stochastic nature and thus can present varying performance levels, for each fold we performed 10 randomly initialized BP runs and computed their 2x2 confusion matrices that are then accumulated to compute the overall confusion matrix (CM). Each CM per fold is then accumulated to obtain the final CM that can be used to calculate the average performance metrics for an overall anomaly detection performance, i.e., classification accuracy (*Acc*), sensitivity (*Sen*), specificity (*Spe*), and positive predictivity (*Ppr*). The definitions of these standard performance metrics using the hit/miss counters obtained from the CM elements such as true positive (*TP*), true negative (*TN*), false positive (*FP*), and false negative (*FN*), are as follows: *Accuracy* is the ratio of the number of correctly classified beats to the total number of beats, $Acc = (TP+TN)/(TP+TN+FP+FN)$; *Sensitivity* (Recall) is the rate of correctly classified A beats among all A beats, $Sen = TP/(TP+FN)$; *Specificity* is the rate of correctly classified N beats among all N beats, $Spe = TN/(TN+FP)$; and *Positive Predictivity* (Precision) is the rate of correctly classified A beats in all beats classified as A, $Ppr = TP/(TP+FP)$.

For all experiments we employ an early stopping training procedure with both setting the maximum number of BP iterations to 50 and the minimum training error to 8% to prevent over-fitting. We use Pc =5 iterations and initially set the learning factor, ε, as $10^{-3}$. We applied a global adaptation during each BP iteration, i.e., if the train MSE decreases in the current iteration we slightly increase ε by 5%; otherwise, we reduce it by 30%, for the next iteration.

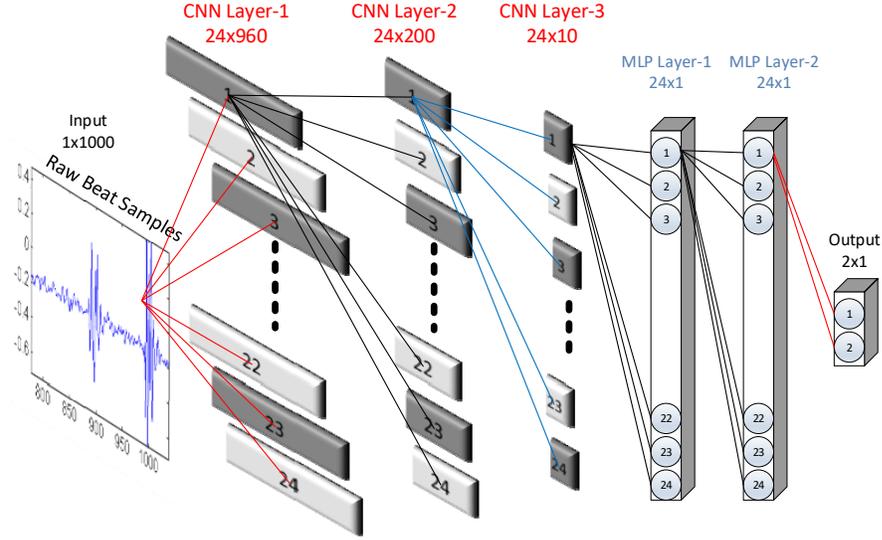

**Figure 6: 1D CNN configuration with 3 CNN and 2 MLP layers.**

### B. Results and Comparative Evaluations

Figure 7 presents two typical examples where the data purification block separated the beats that are classified as N with a high CL (CL(N) > 50%) in abnormal PCG records and eventually dropped out of the BP training until the next periodic check. In the figure N and A beats are shown in the top and bottom sub-plots, respectively. It is clear that there are certain morphological and structural differences among the differentiated beats, i.e., the A beats seem noisier with a higher variance and there are apparent differences among the S1 and especially S2 peaks. The experimental results with the data purification block have shown an average accuracy improvement around 1.8% on the test set and therefore, we can conclude that the proposed method can reduce the confusion caused by the potential N beats in abnormal records.

The comparative evaluations are performed over the PCG records in the test partitions of the 4-folds. It is worth mentioning that we used the competing method with minimum adaptation. To be more specific the competing method was designed for PCG quality and anomaly detection while in this work the focus is only on the anomaly detection. The threshold for the majority rule, *Ta*, is varied between [0.1, 0.4]. As discussed earlier, this threshold determines the minimum rate of A beats in a record to classify it as A. As *Ta* gets lower, it is more likely to detect higher number of actual abnormal records and hence the sensitivity (recall) rate would increase; however, as a trade-off, the false alarms, too, would also increase due to variations and noise effects in normal records. This, in turn, reduces the both *Specificity* and *Positive Predictivity* (Precision). Depending on the target application, this threshold can therefore be set in order to maximize or favor either *Sensitivity* or *Specificity-Positive Predictivity*.

Over the HSNR dataset, Figure 8 and Figure 9 show the average *Spe vs. Sen* and *Ppr vs. Sen* plots with varying *Ta* values, respectively. The red circle in both plots indicates the corresponding performance values achieved by the competing method, i.e., *Sen*= 0.8967 , *Spe*= 0.8689 and *Ppr* = 0.6970. We observed that the proposed approach achieves superior performance levels with respect to all three metrics for a certain range of *Ta* values (i.e., 0.2 < *Ta* < 0.3). Alternatively, setting *Ta* < 0.15 yields only around < 0.04 less *Specificity* and *Positive Predictivity* than the competing method; whilst the proposed approach can now achieve close to 0.96 *Sensitivity* level which basically indicates that except only few, almost all abnormal records are successfully detected.

One performance metric, which has not also been considered in the Physionet/CinC competition is the Positive Predictivity (Precision). The range, *0.63 < Ppr < 0.75*, indicates a high level of false alarms (mis-classification of normal records). This is in fact an expected outcome for this problem considering the inter-class similarities and intra-class variations as displayed in some of the examples in Figure 1. The low-SNR dataset is formed from the PCG records with the highest noise variance as the PCG beats of two sample records are shown in Figure 10. It is clear that the entire morphology of a regular PCG beat is severely degraded.

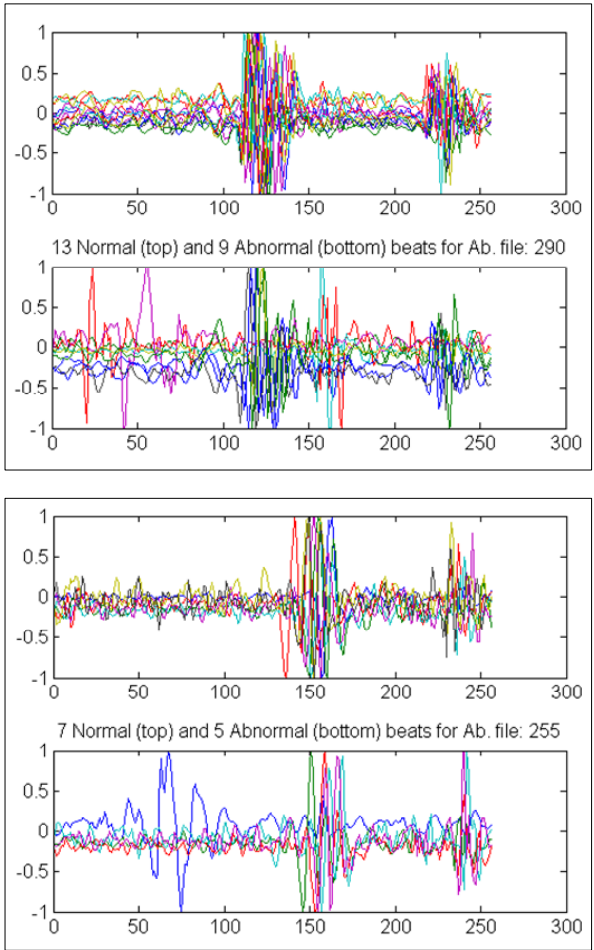

**Figure 7: For two abnormal records with IDs, 290 (top) and 255 (bottom), those beats that are classified as N with a high confidence level ( >50%) are shown at the top sub-plot.**

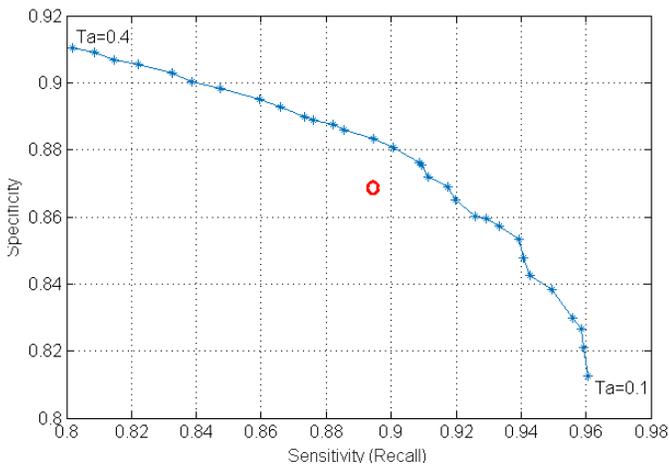

**Figure 8: Average *Specificity* vs. *Sensitivity* (*Recall*) plot with varying *Ta* in the range [0.1, 0.4] for the test partitions of 4-fold cross validation over the high-SNR dataset. The red circle represents the performance of Zabihi et al. (2016).**

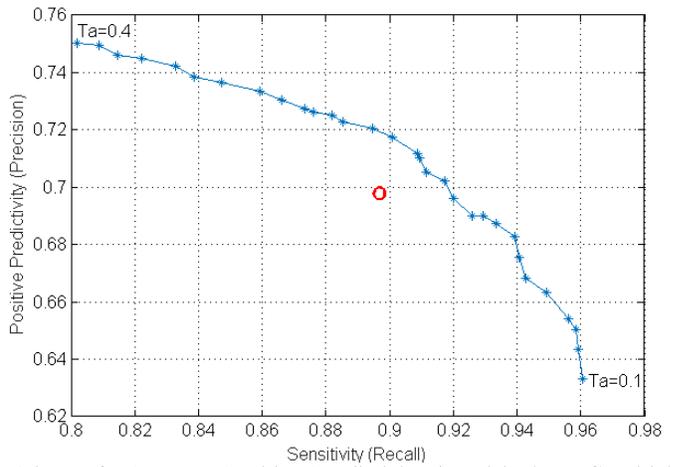

**Figure 9: Average *Positive Predictivity (Precision)* vs. *Sensitivity* (*Recall*) plot with varying *Ta* in the range [0.1, 0.4] for the test partitions of 4-fold cross validation over the high-SNR dataset. The red circle represents the performance of Zabihi et al. (2016).**

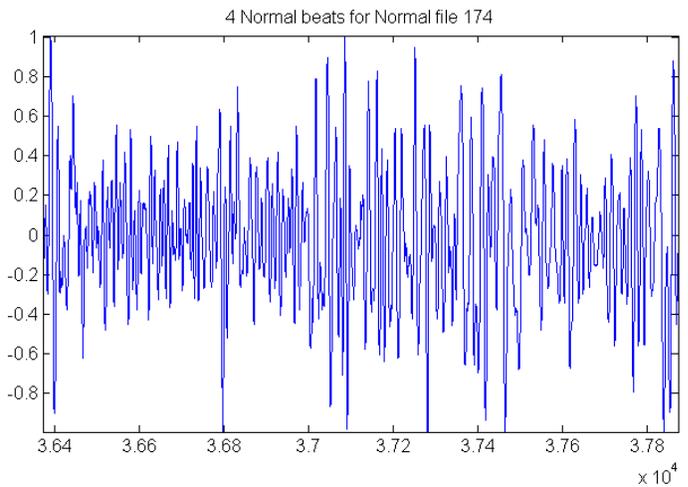

**Figure 10: 4 consecutive N beats from a sample PCG record in low-SNR dataset.**

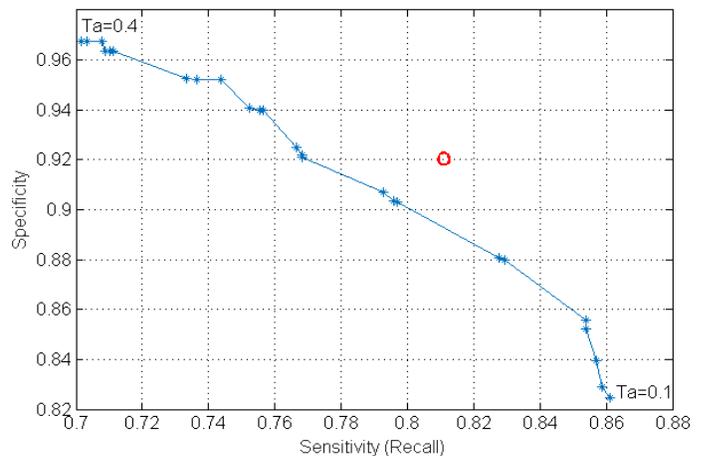

**Figure 11: Average *Specificity* vs. *Sensitivity* (*Recall*) plot with varying *Ta* in the range [0.1, 0.4] for the test partitions of 4-fold cross validation over the low-SNR dataset. The red circle represents the performance of Zabihi et al. (2016).**

We can now evaluate the effect of such (high) noise presence over the proposed approach. For this purpose, Figure 11 and Figure 12 show the average *Spe vs. Sen* and *Ppr vs. Sen* plots with varying *Ta* values over the low-SNR dataset. Once again the red circle in both plots indicates the corresponding performance values achieved by the competing method, i.e., *Sen= 0.8135 , Spe= 0.922* and *Ppr = 0.7795*. Comparing with the results over high-SNR dataset, the high noise level degraded the *Sensitivity* levels of both methods. As an expected outcome, the degradation is worse for the proposed approach since it can only learn from the pattern of a single beat whereas the competing method can use several global features that are based on the joint characteristics and statistical measures of the entire record.

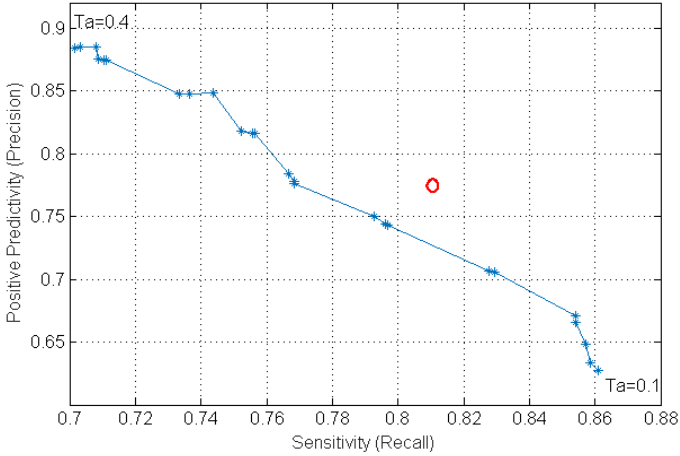

**Figure 12: Average *Positive Predictivity (Precision)* vs. *Sensitivity (Recall)* plot with varying *Ta* in the range [0.1, 0.4] for the test partitions of 4-fold cross validation over the low-SNR dataset. The red circle represents the performance of Zabihi et al. (2016).**

*C. Computational Complexity*

In order to analyse the computational complexity for both FP and BP process, we shall first compute the total number of operations at each 1D CNN layer (ignoring the sub-sampling) and then cumulate them to find the overall computational complexity. During the FP, at a CNN layer, $l$, the number of connections to the previous layer is, $N^{l-1}N^l$. the number of connections to the previous layer is, $N^{l-1}N^l$, there is an individual linear convolution is performed, which is a linear weighted sum. Let $sl^{l-1}$ and $wl^{l-1}$and be the vector sizes of the previous layer output, $s_k^{l-1}$, and kernel (weight), $w_{ki}^{l-1}$, respectively. Ignoring the boundary conditions, a linear convolution consists of $sl^{l-1}(wl^{l-1})^2$ multiplications and $sl^{l-1}$ additions from a single connection. Ignoring the bias addition, the total number of multiplications and additions in the layer $l$ will therefore be:

$$N(mul)^l = N^{l-1}N^l \, sl^{l-1}(wl^{l-1})^2,$$
$$N(add)^l = N^{l-1}N^l \, sl^{l-1} \quad (10)$$

So during FP the total number of multiplications and additions, $T(mul)$, and $T(add)$, on a $L$ CNN layers will be,

$$T_{FP}(mul) = \sum_{l=1}^{L} N^{l-1}N^l \, sl^{l-1}(wl^{l-1})^2,$$
$$T_{FP}(add) = \sum_{l=1}^{L} N^{l-1}N^l \, sl^{l-1} \quad (11)$$

Obviously, $T(add)$ is insignificant compared to $T(mul)$.

During the BP, there are two convolutions performed as expressed in Eqs. (7) and (8). In Eq. (7), a linear convolution between the delta error in the next layer, $\Delta_i^{l+1}$, and the reversed kernel, $rev(w_{ik}^l)$, in the current layer, $l$. Let $xl^l$ be the size of both the input, $x_i^l$, and also its delta error, $\Delta_i^l$, vectors of the $i^{th}$ neuron. The number of connections between the two layers is, $N^{l+1}N^l$ and at each connection, the linear convolution in Eq. (7) consists of $xl^{l+1}(wl^l)^2$ multiplications and $xl^{l+1}$ additions. So, again ignoring the boundary conditions, during a BP iteration, the total number of multiplications and additions due to the first convolution will, therefore, be:

$$T_{BP}^1(mul) = \sum_{l=0}^{L-1} N^{l+1}N^l \, xl^{l+1}(wl^l)^2,$$
$$T_{BP}^1(add) = \sum_{l=0}^{L-1} N^{l+1}N^l \, xl^{l+1} \quad (12)$$

The second convolution in Eq. (8) is between the current layer output, $s_k^l$,, and next layer delta error, $\Delta_i^{l+1}$ where $wl^l = xl^{l+1} - sl^l$. For each connection, the number of additions and multiplications will be, $wl^l$ and $wl^l (xl^{l+1})^2$, respectively. During a BP iteration, the total number of multiplications and additions due to the second convolution will, therefore, be:

$$T_{BP}^2(mul) = \sum_{l=0}^{L-1} N^{l+1}N^l \, wl^l \, (xl^{l+1})^2,$$
$$T_{BP}^2(add) = \sum_{l=0}^{L-1} N^{l+1}N^l \, wl^l \quad (13)$$

So at each BP iteration, the total number of multiplications and additions will be, $(T_{FP}(mul) + T_{BP}^1(mul) + T_{BP}^2(mul))$ and $(T_{FP}(add) + T_{BP}^1(add) + T_{BP}^2(add))$, respectively. Obviously, the latter is insignificant compared to former especially when the kernel size is high. Moreover, both operation complexities are proportional to the total number of connections between two consecutive layers, which are the multiplication of the number of neurons at each layer. Finally, the computational complexity analysis of MLPs is well known (e.g., see Serpen et al., 2014) and it is quite negligible in the current implementation since only a scalar (weight) multiplication and an addition are performed for each connection.

The implementation of the adaptive 1D CNN is performed using C++ over MS Visual Studio 2013 in 64bit. This is a non-GPU implementation; however, Intel ® OpenMP API is used to obtain multiprocessing with a shared memory. We used a 48-core Workstation with 128Gb memory for the training of the 1D CNN. One expects 48x speed improvement compared to a single-CPU implementation; however, the speed improvement was observed to be around 36x to 39x.

The average time for 1 BP iteration (1 consecutive FP and BP) per PCG beat was about 0.57 msec. A BP run typically takes 5 to 30 iterations over the training partition of the high-SNR dataset, which contains around 30000 PCG beats overall. Therefore, the average time for training the 1D CNN classifier was around, 30000*30*0.57msec = 513 seconds (less than 9 minutes) which is a negligible time since this is a one-time (offline) operation.

Besides this insignificant time complexity for training, another crucial advantage of the proposed approach over the competing method is its elegant computational efficiency for the (online) beat classification, which is the main process of heart health monitoring. The average time for a single PCG beat classification is only 0.29 msec. Since the human heart beats 1-4 times in a second, the classification of the 1-4 PCG beats acquired in a real-time system will take an insignificant time which is a mere fraction of a second. For instance for 1 beat/second, this indicates more than 3000x higher speed than a real-time requirement and thus it is practically an instantaneous anomaly detection that can be adapted to even low-power portable PCG monitors in a cost-effective way.

## V. Conclusions

Recently the *PhysioNet (CinC) Challenge* set the state-of-the-art for anomaly detection in PCG records. The best anomaly detection method can classify each PCG record as a whole using a wide range of features over an ensemble of classifiers. In this study our aim is to set a new frontier solution which can achieve a superior detection performance. As additional but equally important objectives that have not been tackled in the challenge, the proposed systematic approach is designed to provide a real-time solution whilst reducing the false alarms. Under the condition that a reasonable signal quality (SNR) is present, the experimental results show that all the objectives have been achieved and furthermore the proposed approach can conveniently be used in portable, low-power PCG monitors or acquisition devices for automatic and real-time anomaly detection. This is a unique advantage of the proposed approach over the best method (Zabihi et al., 2016) of the Physionet challenge, which can only classify an entire PCG record as a whole. On the other hand, for PCG file classification, the unique parameter, *Ta*, for majority rule decision can be set in order to favor either *Sensitivity* or *Specificity – Positive Predictivity* (or both) depending on the application requirements. For instance, such a control mechanism can be desirable for those applications where the computerized results will be verified by a human expert and thus *Ta* can be set to a sufficiently low value, i.e., *Ta*=0.1, so that the *Sensitivity* is maximized (e.g., *Sen* > 96%) in order to accurately detect almost *all* abnormal records in the dataset. This is another crucial advantage over the competing method as it basically lacks such a control mechanism.

In this study we also investigated the fundamental drawback of such global and automated solutions including the one proposed in this paper. This is a well-known fact for ECG classification that has been reported by several recent studies (Jiang et al., 2017; Ince et al., 2009; Kiranyaz et al., 2013; Llamedo et al., 2012; Li et al., 1995; Mar et al., 2011). Similar to ECG, PCG, too, shows a wide range of variations among patients as some examples are shown in Figure 1. In other words, PCG also exhibits patient-specific patterns many of which can conflict with other patient's data during classification. This drawback is also the main reason why further significant performance improvement may no longer be viable with such global approaches. As a result, like in the state-of-the-art ECG studies, we believe that only the patient-specific or personalized PCG anomaly detection approaches can further improve this frontier in a significant way. This will be the topic of our future work.